
Theoretical Study of Friction: A Case of One-Dimensional
Clean Surfaces.

Hiroshi Matsukawa and Hidetoshi Fukuyama+

Department of Physics, Faculty of Science, Nara Women's
University,

Nara 630, Japan

Department of Physics, Faculty of Science, Osaka
University,

1-16, Machikaneyama-cho, Toyonaka, Osaka 560, Japan*
+Department of Physics, Faculty of Science, University of
Tokyo,

7-3-1, Hongo, Bunkyo-ku, Tokyo 113, Japan

(Received 30 November 1993)
Abstract
	A new method has been proposed to evaluate the
frictional force in the stationary state.  This method is
applied to the 1-dimensional model of clean surfaces.  The
kinetic frictional force is seen to depend on velocity in
general, but the dependence becomes weaker as the maximum
static frictional force increases and in the limiting case
the kinetic friction gets only weakly dependent on velocity
as described by one of the laws of friction.  It is also
shown that there is a phase transition between state with
vanishing maximum static frictional force and that with
finite one.  The role of randomness at the interface and
the relation to the impurity pinning of the sliding
Charge-Density-Wave are discussed.


*Present Address
e-mail address: hiro@ena.wani.osaka-u.ac.jp
To appear in Phys.Rev.B.

1. Introduction

The study of friction has been carried out for a long
time1.  Recently the developments of new technologies
enable us to investigate the friction in a situation which
was not considered before2~10.  It is well known that the
three laws of friction hold well in a usual situation1:
(i) The frictional force does not depend on an apparent
area of contact surfaces.  (ii) The frictional force is
proportional to the normal load, where the proportionality
constant is called the coefficient of friction.  (iii) The
kinetic frictional force does not depend on the relative
velocity of contact surfaces and is less than the maximum
static frictional force.  It is also known that there are
cases where these laws of friction no longer hold.  The
limitation of the laws and the possible new laws which hold
beyond such limitation have, however, not been clarified to
the best of our knowledge.  Actually even though there
exists attempts of the theoretical explanation on the laws
of static friction as discussed below, the concrete
explanation of the laws of kinetic friction is lacking.
This paper is our first step to study these basic questions
on friction theoretically.

Theoretical explanation of the laws of static friction
have been as follows1.  Due to the surface roughness the
area of actual contacting points is much less than the
apparent contact area and the pressure there reaches the
yield pressure.  Then the total area of actual contact
points is proportional to the normal load.  The two
surfaces adhere to each other by intermolecular forces at
the contact.  The maximum static frictional force is equal
to the shear strength of the contact times the total area
of contacting points and is then proportional to the normal
load.
This view can not be applied to the clean surface without
surface randomness6.  For example, as the interatomic force
between two bodies works among all atoms at surfaces, the
frictional force is expected to survive even for the
vanishing normal load and to be  proportional to the
contacting area and then the first and second laws of
friction will no longer hold.

Recently Hirano and Shinjo6-8 pointed out another peculiar
feature of the friction of clean surfaces. That is the
frictional transition, which is the phase transition
between states with and without finite maximum static
frictional force.  In the simplest case of 1-dimension(d)
where the atoms of the one body are fixed, the model
reduces to the Frenkel-Kontrova model11.  In that model
such a phase transition is known to exist when the ratio of
mean atomic distances of two bodies is irrational, i.e., in
the incommensurate case.  This transition is called Aubry's
breaking of analyticity transition12.  When the strength of
interatomic potential is less than a critical value the
spatial configuration of atoms in the ground state is
smooth and the maximum static frictional force vanishes.
Above the critical value the configuration has
discommensurate structure and the maximum static frictional
force is finite.  Hirano and Shinjo have claimed the
existence of such a frictional transition even in a 3d
model where the atoms of the lower body are fixed, and that
the static frictional force can vanish between pure metals
with incommensurate clean surfaces, e.g., (111)  and (110)
surfaces of a-iron6,8.

All of these studies are, however, based on the model where
atoms in one of the interfaces have held fixed.  Nothing is
known about the frictional transition in a more realistic
model where both atoms can relax.  The behavior of kinetic
friction has also not been explored yet.

Deferring the detailed studies on rough surfaces until
future publications, we study in this paper the static and
kinetic frictional forces of the 1d model of clean
surfaces, where atoms in both sides of the interface can
relax.  In the theoretical study of friction, it is not
clear even what should be calculated as the frictional
force and a method of calculation has not been established.
We first propose a new method to calculate the kinetic
frictional force in a stationary state and then apply it to
1d model of clean surfaces.  We derive the explicit
velocity dependence of the kinetic friction of the present
microscopic model.  It is seen that the velocity dependence
is appreciable in general.  This dependence becomes weaker
as the maximum static frictional force increases and
eventually the kinetic friction gets almost velocity
independent as described by the third law of friction.  At
the same time we also find the frictional transition in the
present model.

This paper is organized as follows.  In 2 we address
ourselves to the method to calculate the frictional force.
3 we define the 1d model of clean surfaces and discuss the
relationship of the present model with the Frenkel-Kontrova
model.  Then the numerical results based on the model are
presented.  Finally in .4 we summarize the present results
and discuss the effects of randomness and the relation to
the impurity pinning of the sliding Charge-Density-Wave.
\end